\documentclass{emulateapj}
\usepackage{graphicx}

\slugcomment{Submitted for publication in the AJ}

\shorttitle{Radial Velocity Self-Improvement}
\shortauthors{Allende Prieto}

\begin{document}

\title{Velocities from Cross-Correlation: A Guide for Self-Improvement}

\author{Carlos Allende Prieto}
\affil{McDonald Observatory and Department of Astronomy, University of Texas,
    Austin, TX 78712}
\email{callende@astro.as.utexas.edu}

\begin{abstract}

The measurement of Doppler velocity shifts in spectra is  
a ubiquitous theme in astronomy, usually handled by computing 
the cross-correlation of the signals, and finding the location 
of its maximum. This paper addresses the problem of the determination 
of wavelength or velocity shifts among multiple spectra of the same, 
or very similar,  objects. We implement the classical cross-correlation 
method and experiment with several simple models to determine the 
location of the maximum of the cross-correlation function.
We propose a new technique, {\it self-improvement},
to refine the derived solutions
by requiring that the relative velocity for any given pair of spectra is
consistent with all others. By exploiting all available information, 
spectroscopic surveys involving large numbers of similar objects may improve
their precision significantly. 
As an example, we simulate the analysis of a survey of 
G-type stars with the SDSS instrumentation. Applying 
{\it self-improvement}  refines relative radial velocities 
by more than 50\% at low signal-to-noise ratio. The concept  
is equally applicable to the problem of combining a series 
of spectroscopic observations of the  same object, each with a different
Doppler velocity or instrument-related 
offset, into a single spectrum with an enhanced signal-to-noise ratio.

\end{abstract}


\section{Introduction}

The application of cross-correlation techniques to measure velocity shifts
has a long history (Simkin 1972, 1974; Lacy 1977; Tonry \& Davis 1979), and
%Nordstr\"om et al. 1994; Kurtz \& Mink 1998), and
with the advent of massive digital spectroscopic surveys of
galaxies and stars, the subject has renewed interest. The
recently completed Sloan Digital Sky Survey (SDSS) has collected spectra
for more than 600,000 galaxies and 90,000 quasars 
(Adelman-McCarthy et al. 2007, York et al. 2000). 
The SDSS has also obtained spectra for about 200,000 galactic stars, 
and it is now being extended at lower galactic latitudes
by SEGUE with at least as many spectra 
(Rockosi 2005, Yanny 2005). Another ongoing galactic survey, RAVE, 
is expected to collect high-resolution spectra for a million stars by 2011 
(Steinmetz et al. 2006), and the plans for the GAIA satellite
include measuring radial velocities for 10$^8$ stars by 2020 
(Katz et al. 2004). Extracting the maximum 
possible information from these spectroscopic surveys requires
carefully designed strategies.

Cross-correlation has been the target of numerous 
developments in recent years 
(see, e.g., Mazeh \& Zucker 1994,  Statler 1995, 
Torres, Latham \& Stefanik 2007, Zucker 2003), 
but  several practical aspects of its implementation would benefit from 
further research. These include the selection 
of templates (e.g., observed vs. synthetic libraries), 
how to combine measurements from multiple templates, 
the method to determine the maximum of the cross-correlation 
function, data filtering, and error determination. 
Some of these issues are briefly addressed in this paper,
but our focus is on how the requirement of coherence
among all entries in a radial velocity 
data base can be used to improve the original measurements.
A different but plausible approach  has been recently 
proposed by Zucker \& Mazeh (2006).

The Doppler shifts of targets in a spectroscopic survey
are determined one at a time. Each object's projected velocity 
is measured independently, not counting 
a possible common set of cross-correlation templates. 
For a given template, 
from any pair of (projected) velocity measurements, we can derive 
a relative velocity between the two objects involved. However,
that figure will likely be numerically different from the
value inferred from the direct cross-correlation between 
their spectra, even if the two objects are of the same class. 
In this paper, we argue that it is possible to improve the 
original determinations by imposing consistency among
all available measurements. Our discussion is oriented to the case 
of a homogeneous sample: multiple observations of the same  
or similar objects.

In the following section I introduce 
cross-correlation, with a brief discussion about error evaluation.
Section \ref{basic} presents the notion of {\it self-improvement} and 
Section \ref{general} extends the method  
to the more realistic scenario in which the spectra in
a given data set have varying signal-to-noise ratios.
In \S \ref{sdss} we explore an application of the proposed technique 
involving low-resolution spectra, 
concluding the paper with a brief discussion and 
reflections about future work.

\section{Cross-correlation analysis}
\label{xcorr}

The most popular procedure for deriving relative velocities between a
stellar spectrum and a template is the 
cross-correlation method (Tonry \& Davis 1979).
This technique makes use of all the available information in the two 
spectra, and has proven to be far superior than
simply comparing the Doppler shifts between the central wavelengths of 
 lines when the signal-to-noise ratio is low. 
The cross-correlation of two  arrays (or spectra)
{\bf T} and {\bf S}  is defined as a new array {\bf C} 
\begin{equation}
 C_i = \sum_{k} T_k S_{k+i}.
\label{Ci}
\end{equation}
\noindent 
If the spectrum {\bf T} is identical to {\bf S}, but shifted
by an integer number of pixels $p$, the maximum value in the array {\bf C} 
will correspond to its element $i=p$. 
Cross-correlation can be similarly
used to measure shifts that correspond to non-integer numbers. 
In this case, finding the location of the maximum value of 
the cross-correlation function
can be performed with a vast choice of algorithms.

The most straightforward procedure to estimate realistic 
uncertainties involves an accurate noise model and Monte-Carlo simulations,
and that is the method we use in Section \ref{sdss}. 
We employ Gaussians and low-order polynomials to model
the peak of the cross-correlation function. For these simple models,
implemented in a companion IDL code, 
it is possible to derive analytical approximations that relate
the uncertainty in the location of the maximum of the cross-correlation
function to the covariance matrix [U$_{ij}$]. 

\section{Velocity self-improvement}
\label{basic}

Digital cross-correlation, introduced in Section \ref{xcorr},
is commonly employed to derive Doppler radial velocities 
between two spectra. The discussion in
this section is, nonetheless, more general, and deals with the 
statistical improvement of a set of relative velocity measurements.

If three spectra of the same object are available and we refer to the
relative radial velocity  between the first two as $V_{12}$,
an alternative estimate of $V_{12}$ can be obtained by
combining the other relative velocity measurements, $V_{13} - V_{23}$.
Assuming uniform uncertainties, the error-weighted average of 
the two values is $V'_{12}  = (2 V_{12} + V_{13} - V_{23})/3$. 
For a set of $n$ spectra, we can obtain an improved relative radial velocity 
determination between the pair $ij$ by generalizing this expression
\begin{equation}
V'_{ij} = \frac{2 V_{ij}}{n} + \sum_{k=1, \\ k \neq i, k \neq j}^n
\frac{V_{ik} - V_{jk}}{n}.
\label{symmetry}
\end{equation}

It can be seen from Eq. \ref{Ci} that the correlation
of {\bf T} and {\bf S} is equal to the reverse of the correlation
between {\bf S} and {\bf T}. Thus, when the relative velocities
between two spectra is derived from cross-correlation and the 
spectra have a common sampling, it will be satisfied that $V_{ij}= -V_{ji}$, 
but this will not be true  in general. For example, 
if we are dealing with grating 
spectroscopy in air, changes in the refraction index with time may alter the
wavelength scale and the spectral range covered by any particular
pixel, requiring interpolation.
If our choice is to interpolate the second
spectrum ({\bf S}) to the scale of the first ({\bf T}), this may 
introduce a difference between $V_{ij}$ and $-V_{ji}$ due to different
interpolation errors. We can accommodate the general case by writing
\begin{eqnarray}
V'_{ij} & =  & \frac{1}{n} 
\left(V_{ij} - V_{ji} \right) +   \nonumber \\
&  & \frac{1}{2n} \sum_{k=1, k \neq i, k \neq j}^n 
V_{ik} - V_{ki} + V_{kj} - V_{jk}.  
\label{vprime}
\end{eqnarray}
\noindent Note that this 
definition ensures that $V'_{ij} = -V'_{ji}$, and $V'_{ii}=0$.

If the quality of the spectra is uniform, and all measured radial
velocities $V_{ij}$ have independent uncertainties of the same size 
$\sigma \equiv \sigma_{ij}$, the primed values would have an 
uncertainty $\sigma/\sqrt{n}$. Despite $V_{ij}$ may be numerically different
from $-V_{ji}$, $\sigma_{ij}$ will be highly correlated with 
$\sigma_{ji}$, and thus the uncertainty in the primed velocities will 
not be reduced that fast. In addition, all $V_{ij}$ are also correlated with
all $V_{ik}$, 
driving the improvement farther away from the ideal $1/\sqrt{n}$ behavior.
We can expect that after a sufficient number of 
spectra are included, either random errors will shrink 
below the 
systematic ones or all the available information will already be extracted,
and no further improvement will be achieved.

\section{General case of multiple spectra with different signal-to-noise
ratios}
\label{general}

The case addressed in Section \ref{basic} corresponds to a set of
spectra of the same quality. If the uncertainties
in the measured relative radial velocities differ significantly among 
pairs of spectra, 
Eq. \ref{vprime} can be generalized by using a weighted average
\begin{eqnarray}
V'_{ij} = 
\frac{1}{1/\omega_{ij}^2  + \displaystyle \sum_{k \neq i,j} 1/\Omega_{ijk}^2}
\times  ~~~~~~~~~~~~~~~~~~~~~~~~~~~~~~~ \nonumber \\
        ~~~~~~~~~~~~~~~~~~~~~~~~~~~~~~~ \nonumber \\
\displaystyle \bigg\{ V_{ij}/\sigma_{ij}^2 - V_{ji}/\sigma_{ji}^2 + %\right. 
        ~~~~~~~~~~~~~~~~~~~~~~~~~~~~~~~ \nonumber \\
        ~~~~~~~~~~~~~~~~~~~~~~~~~~~~~~~ \nonumber \\
\left. \displaystyle \sum_{k \neq i,j}
\frac{
  \omega_{ik}^2 \left( V_{ik}/\sigma_{ik}^2 - V_{ki}/\sigma_{ki}^2 \right) 
- \omega_{jk}^2 \left( V_{jk}/\sigma_{jk}^2 - V_{kj}/\sigma_{kj}^2 \right)
}{\Omega_{ijk}^2}
\right\} \nonumber \\
\label{vprimegen}
\end{eqnarray}
\noindent where 
\begin{eqnarray}
\sigma_{ij}  \equiv \sigma(V_{ij}), \nonumber \\
\nonumber \\
\displaystyle
\frac{1}{\omega_{ij}^2}  = \frac{1}{\sigma_{ij}^2} + \frac{1}{\sigma_{ji}^2}, 
\nonumber \\
\nonumber \\
\Omega_{ijk}^2  = \omega_{ik}^2 + \omega_{kj}^2, 
\end{eqnarray}
\noindent and the uncertainty is
\begin{equation}
\sigma(V'_{ij}) = 
\left(
\frac{1}{\omega_{ij}^2} + 
\displaystyle \sum_{k \neq i,j} \frac{1}{\Omega_{ijk}^2}
\right)^{-1/2}.
\end{equation}

In the common case in which $V_{ij} = -V_{ij}$, the counterpart
of Eq. \ref{symmetry} for dealing with spectra of varying signal-to-noise
ratios reduces to
\begin{equation}
V'_{ij} = 2 \sigma^2 (V'_{ij})
	\left(
\frac{V_{ij}}{\sigma_{ij}^2}  + \displaystyle \sum_{k \neq i,j}
\frac{V_{ik} - V_{jk}}{\sigma_{ik}^2 + \sigma_{jk}^2}
\right)
\label{vprime2}
\end{equation}
\noindent where
\begin{equation}
\frac{1}{\sigma^2(V'_{ij})} = 
\frac{2}{\sigma_{ij}^2} 
+ \displaystyle \sum_{k \neq i,j}
\frac{2}{\sigma_{ik}^2 + \sigma_{jk}^2}.
\end{equation}

In the next section we use simulated spectra for a  case study: multiple
observations of the same object or  massive surveys
involving large numbers of very similar objects at intermediate 
spectral resolution.

\section{Example: An SDSS-like survey of G-type stars}
\label{sdss}

The SDSS spectrographs deliver a resolving power of 
$R \equiv \lambda/FWHM \simeq 2000$, over the range 381--910 nm. 
These two fiber-fed instruments are attached to a dedicated 
2.5m telescope at Apache Point Observatory (Gunn et al. 2006).
Each spectrograph can obtain spectra for 640 targets simultaneously.
As a result of a fixed exposure time in SDSS spectroscopic
observations, the flux in a stellar spectrum at a 
reference wavelength of 500 nm, $f_{500}$, correlates well 
with the $g$ magnitude of the star and with the signal-to-noise ratio 
at 500 nm ($S/N_{500}$). 
On average, we find $S/N_{500} = 25$ at $g \simeq 17.85$ mag.
To build a realistic noise model, we used  the fluxes and 
uncertainties for 10,000 spectra  publicly released as part of DR2 
(Abazajian et al. 2004)
to derive, by least-squares fitting, a polynomial approximation. 
When $f_{500}$ is expressed in erg cm$^2$ s$^{-1}$ \AA$^{-1}$, 
which are the units used in the SDSS data base, we can write 
\begin{eqnarray}
S/N_{500} \simeq -167386.14 -41737.014 {\mathcal L}
		 & -3884.2333 {\mathcal L}^2 \nonumber \\
		-160.00277 {\mathcal L}^3
		-2.4627508 {\mathcal L}^4, & 
\label{snr500}		
\end{eqnarray} 
\noindent where ${\mathcal L} = \log_{10} f_{500}$.
This relationship holds in the range $-16.5 < {\mathcal L} < -14.5$.

The uncertainties in the SDSS fluxes for stars -- mostly 
relatively bright calibration sources-- are not dominated by photon
noise, but by a {\it floor} noise level of 2--3\% associated with 
a combination
of effects, including imperfect flat-fielding and scattered light corrections.
Errors are highly variable with wavelength, but  
the noise at any given wavelength depends linearly on the signal.
Based on the same set of SDSS spectra used for Eq. \ref{snr500}, 
we determine the coefficients  in the relation 
\begin{equation}
\sigma_{\lambda} = a_{\lambda} + b_{\lambda} f_{\lambda},
\label{snr}
\end{equation}
which we use here for our numerical experiments. 
%These coefficients are provided in Table 1.
For a given choice of $S/N_{500}$, we interpolate the table of
coefficients $a_{\lambda}$ and $b_{\lambda}$ derived from SDSS data, and by 
inverting Eq. \ref{snr500} we derive the flux at 500 nm. Finally, we scale the
spectrum fluxes and calculate the expected errors at all wavelengths
using Eq. \ref{snr}. 
Gaussian noise is introduced for each pixel position, 
according to the appropriate error, simulating multiple 
observations of the same star to create an entire library of spectra.

\begin{figure*}[ht!]
\centering
\includegraphics[width=10.cm,angle=90]{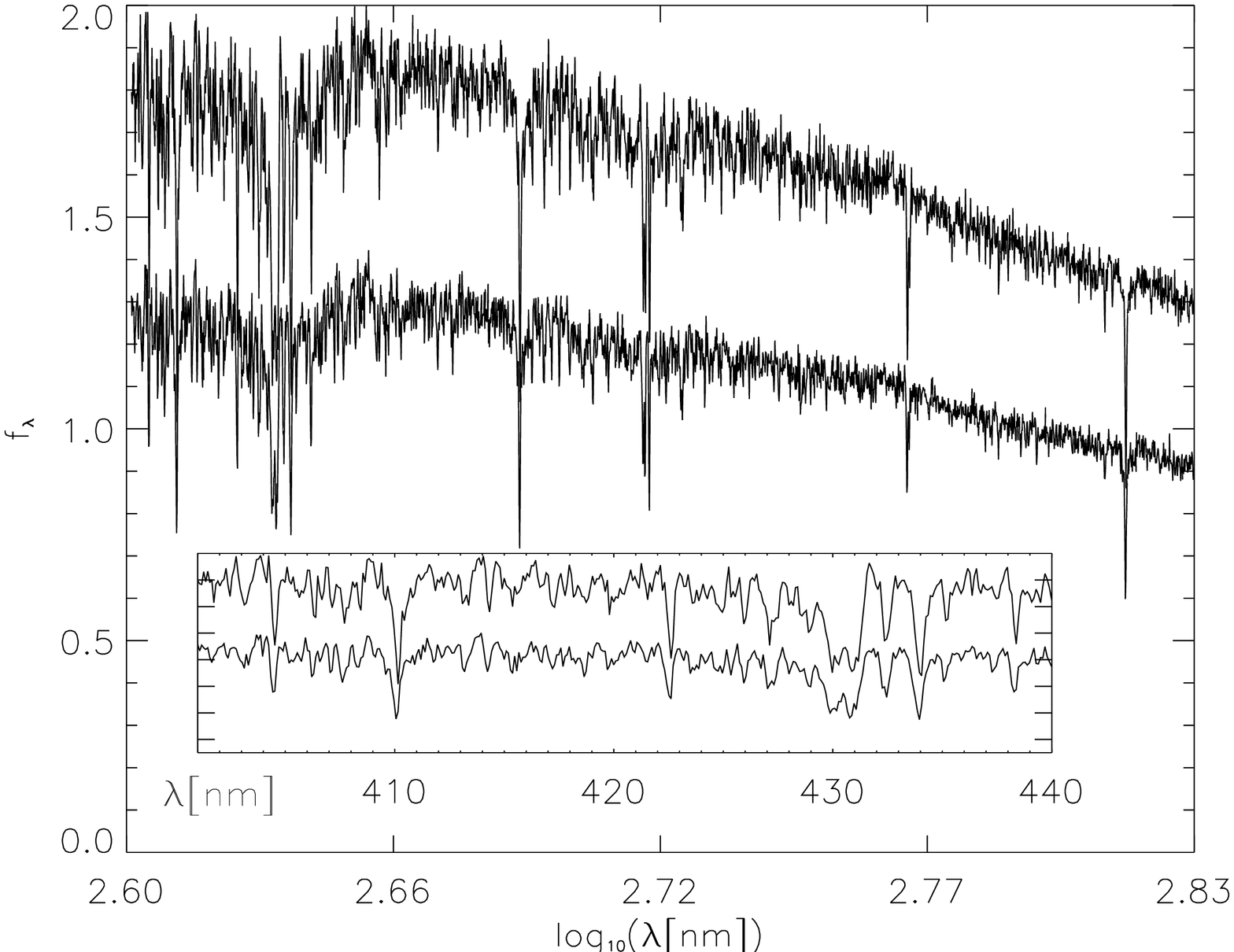}
\includegraphics[width=5.0cm,angle=0]{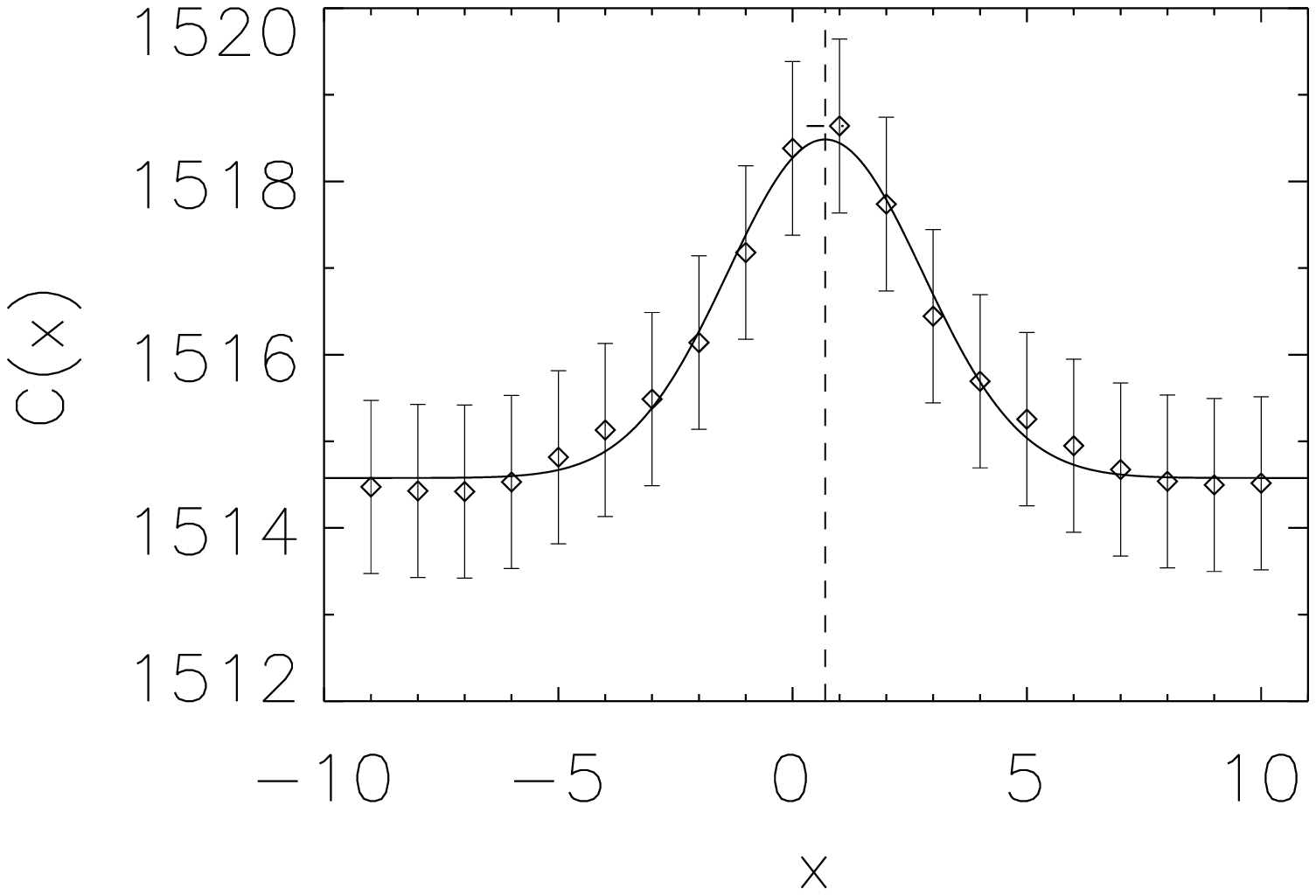}
\includegraphics[width=5.0cm,angle=0]{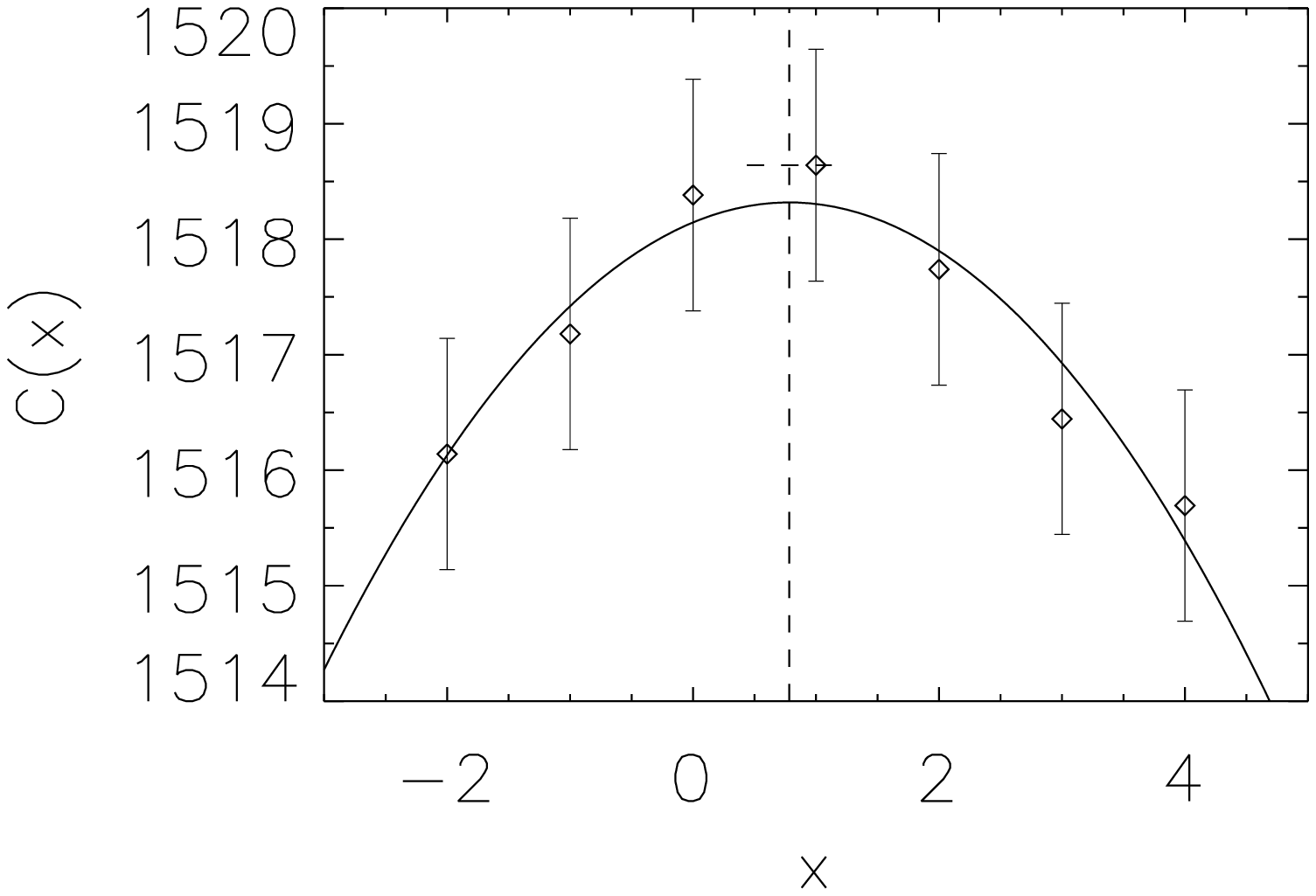}
\includegraphics[width=5.0cm,angle=0]{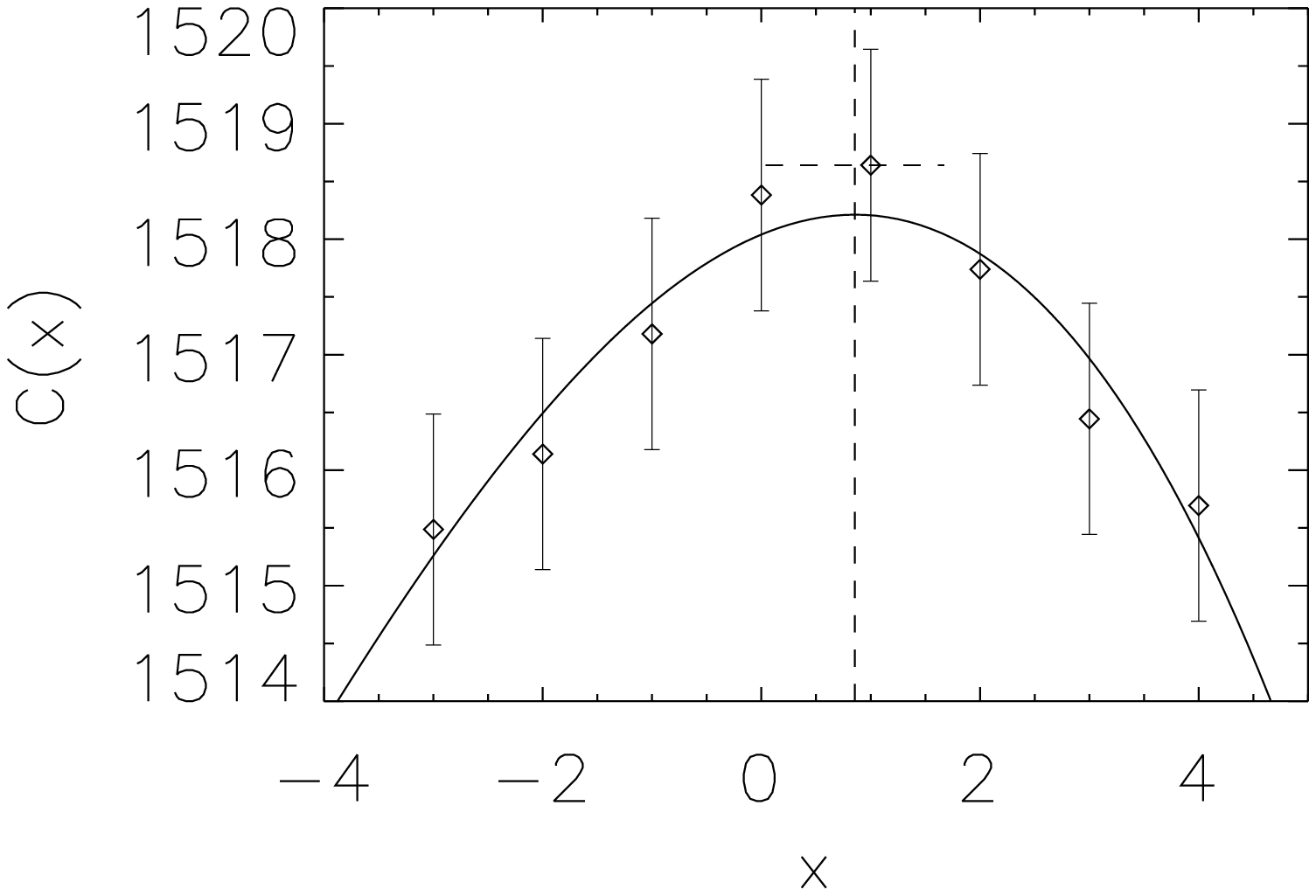}
\protect\caption[ ]{
Two sample spectra used in the tests described in \S \ref{sdss1}.
They correspond to the same original spectrum, shifted by an
arbitrary radial velocity, and with added Gaussian noise so that
$S/N_{500}=50$. (The fluxes have been also arbitrarily scaled for display). 
The inset in the upper panel is a close-up of the blue
part of the spectrum. The lower panels show the cross-correlation
vector and illustrate the determination of the maximum by using
Gaussian, quadratic, and  cubic  models with $N_f = 20, 7$, and 8
data points, respectively. The  dashed vertical lines show the location of
the maxima, and the horizontal dashed lines indicate the estimated
uncertainties.
}
\label{xcf}
\end{figure*}

We employed a spectrum of HD 245, a nearby G2 star\footnote{
Metallicity ([Fe/H]$ \simeq -0.7$), surface gravity ($\log g \simeq 3.7$),
and kinematics, make this object a prototypical thick-disk turn-off star.
}, to produce spectra that resemble SDSS observations 
with various signal-to-noise ratios. 
Radial velocities are also artificially introduced. 
The spectrum of HD 245 used here has a resolving power of 
$R \simeq 10,000$\footnote{
The resolving power of this spectrum 
varies somewhat with wavelength from about $R \simeq 11000$ at 660 nm 
to 7700 at 480 nm. This variation is, however, irrelevant when 
smoothing the data to $R=2000$ as we do in these experiments.
} and is included in the Elodie.3 database 
(Moultaka et al. 2004, Prugniel \& Soubiran 2001). 
As the rest of the library, this spectrum 
was obtained with the 1.9m telescope 
and the Elodie spectrograph at Haute Provence. The original fluxes are resampled 
to $\ln \lambda$, and then smoothed to $R=2000$ by Gaussian 
convolution. The output fluxes are sampled with 12 pixels per 
resolution element. The Doppler shift due to the actual radial velocity of 
HD 245 has already been corrected in the Elodie library. 
New values for the radial velocity in the library of simulated SDSS observations
are drawn from a Normal distribution with a $\sigma=120$ km s$^{-1}$, 
as to approximate the typical range found in F- and G-type 
stellar spectra included in the 
SDSS (mostly thick-disk and halo stars). The wavelength 
scale is then 'Doppler' shifted, changed to vacuum ($\lambda_0$), 
and the spectrum resampled with a step of $10^{-4}$ in $\log_{10} \lambda_0$
(approximately 2.17 pixels per resolution element).
The Elodie spectra only cover the range $400< \lambda_{\rm air} <680$ nm, and
therefore a similar range is finally kept for the SDSS-style files, which
include 2287 pixels.

\subsection{Cross-correlation}
\label{sdss1}

We employed a set of 40 test spectra with $S/N_{500} = 50$,
measuring the relative radial velocities for all possible pairs. 
Fig. \ref{xcf} illustrates two sample spectra and 
their cross-correlation function. To avoid very large or small numbers, the
input arrays are simply divided by their maximum values before 
cross-correlation. We used second and third order
polynomials, as well as a Gaussian to model the 
cross-correlation function and estimate the location of 
its maximum by least-squares fitting. 
The solid line in the lower panels of the figure are the best-fitting models. 
We experimented varying the number of pixels involved in the 
least-squares fittings ($N_f$).

With the sampling used, the measured relative shifts in pixel space ($x$)
correspond to a velocity $V = xc \times 10^{-4}/\log_{10} e$,
where $c$ is the speed of light in vacuum; 
one pixel corresponds to 69 km s$^{-1}$.
%We compare the known relative velocities between all pairs of
%spectra with the values derived from the measurement of
%the location of the cross-correlation peaks. 
We compare the relative velocities between all pairs of spectra
derived from the measurement of
the location of the cross-correlation peaks with the {\it known},
randomly drawn, relative velocities. The average difference 
for the 1600 velocities (40 spectra)
$<\delta V>$ and the rms scatter ($\sigma$) are used to quantify
systematic and random errors, respectively.

Our experiments exhibit no systematic errors 
in the derived velocity when the number of points entering the fit 
$N_f$ was an odd number, i.e., when we use the same number of
data points on each side from the pixel closest to the peak of 
the cross-correlation function.
Modest offsets ($<\delta V>/\sigma \sim 0.02$), however, 
are apparent when fitting polynomials to an
even number of data points, despite we enforce the maximum to be bracketed
by the two central data points. 

Random errors increase sharply with the number of data points involved
in the fittings for the polynomial models, but not for the Gaussian model.
The best results for the polynomials are obtained when the lowest possible 
orders are used. 
Using less than 11 points 
for the Gaussian did not produce reliable results, as there was not enough
information to constrain all the parameters of the model, which includes 
a constant base line.
The best performance $\sigma = 2.8$ km s$^{-1}$ was obtained  
using a second order polynomial
and $N_f=3$. Using a Gaussian model achieved a minimum $\sigma = 4.3$ 
km s$^{-1}$, fairly independent of $N_f$. The third order polynomial provided
the poorest performance, $\sigma = 5.4$ km s$^{-1}$ at best.

The cross-correlation can be computed in Fourier space, taking
advantage of the correlation theorem (Brigham 1974).
This fact is usually exploited to speed up the calculation dramatically, 
as fast Fourier transforms can be calculated with a number of
operations proportional to $N \log_2 N$, compared to $N^2$ required by 
Eq. \ref{Ci}.
Note, however, that for medium-resolution surveys of 
galactic stars, the velocity offsets, limited 
by the galactic escape velocity, usually 
correspond to a limited number of pixels. Therefore, it is
only necessary to compute the values of {\bf C}  
in the vicinity of the center of the array, rendering the timing for a 
direct calculation similar to one performed in transformed 
space\footnote{
In our tests with a Sun Fire V210 (1 CPU) and IDL 6.1, the
timings were similar in pixel or Fourier space:  a
a single cross-correlation ($N=2287$) took  
$2.5-2.6 \times 10^{-3}$ seconds in 
Fourier space (arrays padded to $5^5$ or $2^{12}$), while in pixel space, 
with a lag range restricted to $\pm 15$ pixels ($\pm 1035$ km s$^{-1}$), 
it took $2.2 \times 10^{-3}$ seconds.
}.

\begin{figure}[t!]
\centering
\includegraphics[width=8.4cm,angle=0]{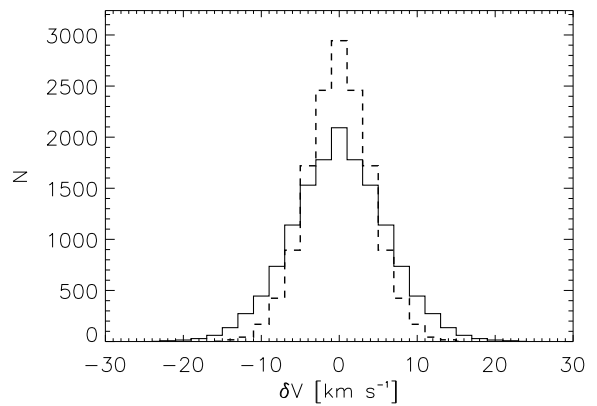}
\protect\caption[ ]{
Distribution of errors in the relative radial velocities derived
from cross-correlation (quadratic model) 
for the case of $n=120$ spectra and 
$S/N_{500} = 25$. The solid line represents the original distribution,
and the dashed line the result after applying self-improvement. 
The error distributions are symmetric because 
the array {\bf V} is antisymmetric.
}
\label{dist}
\end{figure}

\begin{figure*}[ht!]
\centering
~~~~~~~~~~~~~~~~~~~~~~~~~~~
\includegraphics[width=12.cm,angle=90]{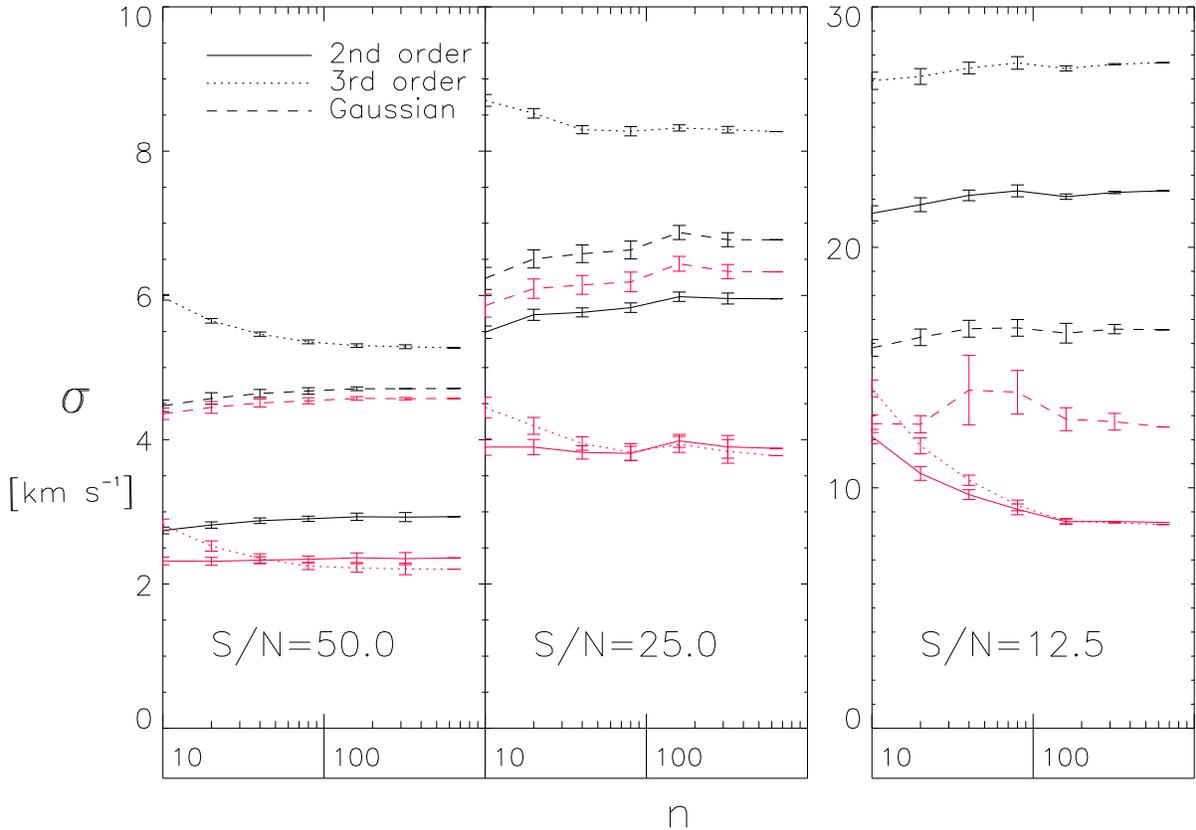}
\protect\caption[ ]{
Estimated $1\sigma$ uncertainties in the cross-correlation between
all pairs in a set of noise-injected spectra of HD 245
as a function of the number of spectra ($n$) for three values of
the signal-to-noise ratio at 500 nm. Three models are considered
to determine the maximum of the cross-correlation function.
The black lines correspond  to the original measurements, 
and the red lines to the final uncertainties after self-improvement. 
}
\label{sigma}
\end{figure*}

\subsection{Self-improvement}
\label{sdss2}

To test the potential of the proposed self-improvement technique
we repeat the same exercise described in \S \ref{sdss1}, but using increasingly
larger datasets including up to 320 spectra, and adopting three
different values for the $S/N$ per pixel at 500 nm: 50, 25, and 12.5.
We calculated the
cross-correlation between all pairs of spectra (matrix {\bf V}), 
and performed quadratic fittings to the 3 central data points, 
cubic polynomial fittings to the central 4 points, and Gaussian fittings 
involving the 11 central points.
We estimated the uncertainties in our measurements
by calculating the rms scatter between the derived and the known
relative velocities for all pairs. Then we applied Eq. \ref{symmetry}
to produce a second set of {\it self-improved} velocities.
(Because the array of wavelengths, $x \propto \log_{10} \lambda_0$, is 
common to all spectra, the matrix {\bf V} is antisymmetric and 
we can use Eq. \ref{symmetry} instead of Eq. \ref{vprime}.)
A first effect of the transformation from {\bf V} to {\bf V'},
is that the systematic offsets described in \S \ref{sdss1} when 
using polynomial fittings with even values of $N_f$ disappear (the
same systematic error takes place for measuring $V_{ij}$ and $V_{ji}$,
 canceling out when computing $V'_{ij}$).

More interesting are the effects on the width of the error
distributions. Fig. \ref{dist} illustrates the 
case when a quadratic model is used  for $n=120$ and $S/N_{500} = 25$.
The solid line represents the original error distribution  and
the dashed line the resulting distribution after self-improvement.
Fig. \ref{sigma} shows the rms scatter as 
 a function of the number of spectra 
for our three values of the $S/N$ ratio at 500 nm. 
The black lines lines show the original results, and the red lines
those obtained after self-improvement. Each panel shows
three sets of lines: solid for the quadratic model, dotted for the
cubic, and dashed for the Gaussian. Extreme outliers at 
$|V|> 2000(S/N_{500})^{-1}$ km s$^{-1}$, if any, were
removed before computing the width of the error distribution ($\sigma$). 
Note the change in the vertical
scale for the case with $S/N_{500}=12.5$. For the experiments 
with $n<80$, several
runs were performed in order to improve the statistics, and the
uncertainty (standard error of the mean) is indicated by the error bars. 
These results are based on the Gaussian random-number generation 
routine included in IDL 6.1, but all the experiments were repeated 
with a second random number generator\footnote{
The F90 module {\tt random.f90} by Alan J. Miller, and in particular the 
normal random number algorithm presented by Leva (1992), coupled
to the intrinsic {\tt random\_number} function included with 
the Sun WorkShop 2.0 F90 compiler.
} and the results were consistent.

As described in \S \ref{sdss1},
the quadratic model performs better on the original velocity 
measurements for $S/N_{500}=50$ and $25$. At the lowest considered
$S/N$ value of 12.5, however, the Gaussian model delivers more
accurate measurements. Self-improvement reduces the errors in all cases. 
Although a second order polynomial fitting works better than
third order for the original measurements, the two models deliver a similar
performance after self-improvement. 
Interestingly, the impact of self improvement is smaller 
on the results from Gaussian fittings than on those
from  polynomial fittings. 
As expected, the errors in the original measurements
are nearly independent of the number of spectra in the test, but 
there is indication that at low signal-to-noise 
the errors after self-improvement for the polynomial models 
decrease as the sample increases in size, until they plateau for  
$n>100$. 

From these experiments, we estimate that the best accuracy attainable 
with the original cross-correlation measurements 
are about 3, 6, and 15 km s$^{-1}$ at $S/N_{500} \simeq 50$, 25, 
and 12.5, respectively. Our results also indicate that 
by applying self-improvement to samples of a few hundred spectra, 
these figures could improve to roughly 2.5, 4, and 9 km s$^{-1}$ 
at $S/N_{500} \simeq 50$, 25, and 12.5, respectively. 

We obtained an independent
estimate of the precision achievable by simply measuring for 320 spectra
the wavelength shift of the core of several strong lines (H$\alpha$, H$\beta$,
H$\gamma$, and H$\delta$) relative to those measured
in the solar spectrum (see Allende Prieto et al. 2006),
concluding that radial velocities can be determined from line 
wavelength shifts
with a $1\sigma$ uncertainty of 3.8 km s$^{-1}$ at $S/N=50$, 
7.2 km s$^{-1}$ at $S/N=25$, and 15.9 km s$^{-1}$ at $S/N=12.5$ 
-- only 10--20\% worse than straight cross-correlation-- but these
absolute measurements cannot take advantage of the self-improvement technique.

Allende Prieto et al. (2006) compared radial velocities determined from
the wavelength shifts of strong lines for 
SDSS DR3 spectra of G and F-type dwarfs with the SDSS pipeline measurements
based on cross-correlation. The derived $1\sigma$ scatter between the
two methods was 12 km s$^{-1}$ or, assuming similar performances, 
a precision of 8.5 km s$^{-1}$ for a given method\footnote{
Note that this figure was determined by fitting a Gaussian profile to the
distribution of differences, and  it will increase
somewhat if a straight rms difference is considered instead.
On the other hand, the spectral types of the stars considered spanned a
broader range (F and G) than in our simulations.
}.
The spectra employed in
their analysis have a $S/N_{500}$ distribution 
approximately linear between 
$S/N_{500}=10$ and 65, with $dN/d(S/N) \sim 7.7$ and with 
mean and median values of 22 and 18, respectively. 
Their result is in line with the expectations 
based on our numerical tests that indicate a potential precision 
of 6--7 km  s$^{-1}$ at $S/N_{500}=25$.
Independent estimates by the SDSS team 
are also consistent with these values (Rockosi 2006; see 
also {\tt www.sdss.org}).

After correcting for
effects such as telescope flexure, the wavelength scale for stellar spectra
in DR5 is accurate to better than 5 km s$^{-1}$ (Adelman-McCarthy et al. 2007). 
This value, derived from the
analysis of repeated observations for a set of standards 
and from bright stars in the old open cluster M67, 
sets an upper limit to the accuracy of 
the radial velocities from SDSS spectra, but  
random errors prevail for $S/N<25$.
Provided no other source of systematic errors is present, 
our tests indicate that 
self-improvement could reduce substantially the typical error bars of  
radial velocities  from low signal-to-noise SDSS observations.

\section{Discussion and conclusions}

This paper deals with the measurement of relative Doppler shifts
among a set of spectra of the same or similar objects. 
If random errors limit the accuracy of the measured relative velocity between 
any two spectra, there is potential
for improvement  by enforcing self-consistency among all
possible pairs.

This situation arises, for example, when 
a set of spectroscopic observations 
of the same object are available and we wish to co-add them to increase
the signal-to-noise ratio.  The spectra may be offset 
due to Doppler velocity offsets or instrumental effects, 
the only difference being that in the former case the spectra should
be sampled uniformly in velocity (or $\log \lambda$) space for  
cross-correlation, while in the latter a different axis 
may be more appropriate.

Another application emerges in the context of surveys 
that involve significant numbers of spectra of 
similar objects. Radial velocities for individual objects 
can be derived using a small set of templates and later {\it self-improved}
by determining the relative velocities among all the survey targets and 
requiring consistency among all measurements. The potential of this technique
is illustrated by simulating spectra for a fictitious survey of G-type
turn-off stars with the SDSS instrumentation. Our simulations show that
applying self-improvement has a significant impact on the potential accuracy
of the determined radial velocities. 
The tests performed dealt with relative 
velocities, but once the measurements are linked to an absolute
scale by introducing a set of well-known radial velocity standards in
the sample, the relative values directly translate into absolute measurements.
The ongoing SEGUE survey includes,
in fact, large numbers of G-type stars, and therefore our results have
practical implications for this project.

The proposed scheme handles naturally the case when multiple
templates are available. Templates and targets 
are not treated differently. Relative velocities are measured for
each possible pair to build $V_{ij}$, and consistency is imposed 
to derive $V'_{ij}$ by using Eqs. \ref{vprime} or \ref{vprimegen}. If,
for example, the templates  have been corrected for their own velocities 
and are the first 10 spectra in the sample, 
the velocity for the $j-$th star ($j>10$)
can be readily obtained as the weighted average of the $V'_{ij}$ elements,
where $i$ runs from 1 to 10. The final velocities would take advantage
of all the available spectra, not just the radial velocity templates,
with differences in signal-to-noise among spectra already 
accounted for automatically.

Very recently, Zucker \& Mazeh (2006) have proposed another approach
with the same goals as the method discussed here. Their procedure
determines the relative velocities of a set of $n$ spectra by
searching for the Doppler shifts that maximize the value of the parameter
$\rho = (\lambda_M -1)/(n-1)$, where $\lambda_M$ is 
the maximum eigenvalue of the correlation matrix
-- a two-dimensional array whose $ij$ element is
is the cross-correlation function between spectra $i$ and $j$. 
Zucker \& Mazeh's algorithm is quite different from the
self-improvement method presented here. It involves finding the set
of velocities that optimally aligns the sample spectra, whereas
self-improvement consists on performing very simple algebraic operations on 
a set of radial velocities that have already been measured. 
Self-improvement is obviously more simple to implement, but a detailed
comparison between the performance of the two algorithms in practical
situations would be very interesting.

This paper also touches on the issue of error determination 
for relative radial velocities derived from cross-correlation, 
and convenient analytical expressions are
implemented in an IDL code available online\footnote{
Together with the
electronic version of this article or from
http://hebe.as.utexas.edu/stools/
}.
We have not addressed many other elements that can potentially
impact the accuracy of Doppler velocities from cross-correlation,
such as systematic errors, filtering, sampling, or template selection.
The vast number of spectra collected by current and planned spectroscopic
surveys should stimulate further thought on these and other issues with
the goal of improving radial velocity determinations. 
There is certainly an abundance of choices that need to be made wisely.

\acknowledgments

This work has 
benefited greatly by input from  David Schlegel, Brian Yanny, and the
referee, Douglas Gies.
Support from NASA (NAG5-13057, NAG5-13147) is thankfully acknowledged.

\end{document}